\documentclass[prx,aps,twocolumn,floats,showpacs,final,groupedaddress]{revtex4-2}
\usepackage{leading}
\usepackage[latin3]{inputenc}
\usepackage[makeroom]{cancel}
\usepackage{graphicx}
\usepackage{amsmath}
\usepackage{amsfonts}
\usepackage{amssymb}

\usepackage{color}
\usepackage[left=2cm,right=2cm,top=2cm,bottom=2cm]{geometry}
\usepackage{graphicx} 
\usepackage{dcolumn}
\usepackage{bm}
\usepackage{simplewick}
\usepackage{array}
\usepackage{appendix}
\usepackage{cancel}
\usepackage[normalem]{ulem} 
\RequirePackage[
   hyperindex,colorlinks,bookmarksnumbered,
   plainpages=true,pdfstartview=FitH]{hyperref}
\hypersetup{linkcolor=blue,urlcolor=blue,citecolor=blue}
\usepackage{hyperref}
\usepackage{mathrsfs}

 \renewcommand{\emph}[1]{\textit{#1}}

\definecolor{darkblue}{rgb}{0,0,0.5}
\definecolor{darkgreen}{rgb}{0,0.5,0}
\definecolor{darkred}{rgb}{.7,0,0}
\definecolor{purple}{rgb}{0.5,0,0.6}
\definecolor{orange}{rgb}{1,0.5,0}
\definecolor{grey}{rgb}{.6,.6,.6}
\definecolor{lightpink}{rgb}{1,0.7,0.75}
\definecolor{pink}{rgb}{1,0.4,0.58}
\definecolor{deeppink}{rgb}{1,0.08,0.58}
\usepackage{upgreek}

\newcommand{\ngg}{N_{\rm g}}

\begin{document}
\title{Tunneling in multi-site mesoscopic quantum Hall circuits}

\author{D. B. Karki}
\affiliation{Department  of  Physics,  Korea  Advanced  Institute  of  Science  and  Technology,  Daejeon  34141,  Korea}

\begin{abstract}
Transport properties of single- and two-site mesoscopic quantum Hall (QH) circuits at high transparencies can be described in terms of the lowest-order backscattering processes, enabling a mapping to the boundary sine-Gordon model. We show that this description breaks down in circuits with four or more sites, where higher-order backscattering processes become relevant and qualitatively modify the low-energy physics, while remaining exactly marginal in three-site geometries. Focusing on the four-site circuit, we derive an effective low-energy theory that captures the resulting interaction-driven physics and reveal the emergence of unique quantum-critical points. In the vicinity of these critical points, we obtain universal conductance and scaling behavior and establish the robustness of the associated non-Fermi liquid physics. We further introduce tunneling in multichannel multi-site QH circuits and propose a promising route for realizing diverse quantum-critical phenomena. We show that a boundary sine-Gordon description can be restored in multichannel multi-site QH circuits by appropriately looping selected edge channels, a procedure that is experimentally feasible. Finally, we analyze the non-equilibrium heating effects relevant to transport measurements in QH circuits. Altogether, our results establish multi-site QH circuits as a versatile and highly controllable platform for simulating interaction-driven quantum critical phenomena.
\end{abstract}

\date{\today}
\maketitle

%%%%%%%%%%%%%%%%%%%%%%%%%%%%%%%%%%%%%%% TITLE %%%%%%%%%%%%%%%%%%%%%%%%%%%%%%%%%%%%
\section{Introduction}
In a typical low-dimensional quantum system, the interplay among strong electron interactions, quantum interference, and different degeneracies generally gives rise to a number of strongly correlated phenomena~\cite{Giamarchi2003}. Manifestations of these effects can also be studied using a well-controllable mesoscopic device, called the single-electron transistor~\cite{sett}. It can be realized as a mesoscopic quantum Hall (QH) circuit consisting of a floating metal grain coupled to one-dimensional QH edge channels~\cite{Matveev_1991,AL,Averin1992}. This metal-semiconductor hybrid structure can exhibit various quantum effects such as Coulomb blockade, the Kondo effect, charge and spin fractionalization, Luttinger liquid behavior, and quantum phase transitions~\cite{Matveev_1991, Flensberg_1993, Matveev1995, Furusaki1995b, Goldhaber_nat(391)_1998,Pierre_2015,Pierre_2018a, Pierre_2018, ssm}. These effects constitute a paradigm in the field of strongly correlated electron physics.

A number of remarkable phenomena have been predicted to emerge in a two-site QH circuit comprising
coupled hybrid metal-semiconductor islands due to the competition between the screening of each individual grain charge and the mediated charge coupling between the two grains~\cite{ccd, KBM}, i.e., the interplay of local and nonlocal correlation effects (see also Refs.~\cite{a1, a2, matveev1996, rok}). Such competition also results in local parafermionic excitations~\cite{ KBM1}. We note that, unlike the Majorana modes (which could be realized in a single-site QH circuit~\cite{EK1, Furusaki1995b}), parafermions are much more complex, since they cannot arise in noninteracting models, being the result of strong correlation effects~\cite{mms}. Therefore, the recent experiment~\cite{ccd} and theoretical predictions~\cite{KBM, KBM1} open up coveted pathways to further understand and manipulate the exotic parafermionic state of matter using a highly tunable nanoelectronic device without topological order. In addition, by adjusting the number of channels connecting the two sites, exotic zero-temperature quantum critical phenomena with diverse critical exponents $\eta\leq 1/2$ can be realized~\cite{dde1}.

One of the promising research directions is to scale up such
hybrid structures to form a lattice that could provide different routes to constructing more complex Fibonacci anyons by leveraging the exotic features of generalized
coupled-impurity models. These multi-site QH circuits can also provide a useful way of understanding the effects of disorder on QH edges due to QH puddle formation~\cite{sssw} (see also Refs.~\cite{dcg, mill}). In addition, they show novel aspects of heat transport~\cite{ kkr, viet,genr, eu1,oliever, es1, para}. It is important to note that owing to the presence of an infinitely large number of electrons in
the metal islands, each grain in the chain, essentially behaves identically to the others. The well-defined QH
edge states implemented via quantum point contacts in the metal-semiconductor hybrid structures also provide unprecedented tuning of the device. Therefore, multi-site QH circuits could be used to simulate different model quantum systems in strongly correlated physics, i.e., they could capture the essential features of a real material~\cite{dggnew}.

It is well understood that the combined effects of a dynamical Coulomb blockade and the integer QH effect generate a form of Luttinger liquid (LL)~\cite{saffi}. A single channel QH circuit consisting of $N$ grains can be described by an equivalent model of an impurity in an LL with the Luttinger parameter $K=1/(N+1)$, or of a fractional QH effect with filling fraction $\nu=K$~\cite{KBM}. As long as there are no tunneling/backscattering events in the system, all the properties of multi-site QH circuits can be achieved simply through the usual circuit analysis~\cite{Slobodeniuk_2013}.
\begin{figure*}[ht!]
\begin{center}
\includegraphics[scale=0.38]{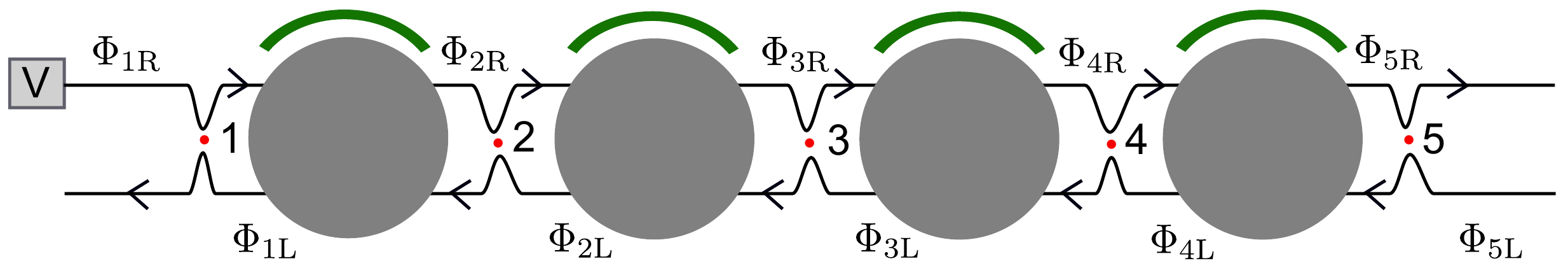}
\caption{Schematic of the four-site quantum Hall circuit consisting of four essentially identical metallic islands connected by five external quantum point contacts (QPCs). Each island is connected to a separate external gate (green metallic plates). The red dots represent the backscattering centers of fully tunable QPCs. The right-moving edge state on the left is biased by a voltage $V$, and the charge current is measured at QPC5. The symbol $\Phi_{\rm jR/L}$ stands for the bosonic field describing the right-/left-moving edge state.}\label{fig1}
\end{center}
\end{figure*}

In the case of a finite backscattering between counter-propagating edge states forming the circuit, the nature of the relevant perturbations depends on the typical value of $\nu$. In the standard problem of tunneling through a weak barrier represented by a potential scatterer $\mathcal{V}(x)$ localized in the vicinity of $x=0$, the relevant boundary perturbations correspond to the process of backscattering of $n$ electrons from one Fermi point to the other by momenta $2k_{\rm F}$~\cite{Kane_1992,ffrs}. These perturbations are represented by the Fourier components of the backscattering potential, i.e.,  $\mathcal{V}(2nk_{\rm F}),\;n=1, 2, \cdots$, and are commonly called as $n$th-order backscattering processes. For all values of $\nu$, the $\mathcal{V}(2 k_{\rm F})$ scattering is always relevant in the renormalization group (RG) sense, and all higher order scatterings that satisfy $n^2\nu<1$ are also relevant perturbation~\cite{Kane_1992}. This shows that only the single-site $N=1$, and two-site $N=2$ QH circuits can fully be described by the lowest-order backscattering processes $\mathcal{V}(2k_{\rm F})$ where mapping to the boundary sine-Gordon model can be exploited in full generality~\cite{Fendleyz,Fendleyw}. For $N=3$, the $\mathcal{V}(4k_{\rm F})$ processes is exactly marginal, and introduce only trivial renormalization effects. Therefore, to study the transport properties of multi-site ($N\geq 4$) QH circuits at large transparencies, it is crucial to also consider the higher-order backscattering processes. The same thing is also true even for $N=1$ and $N=2$ but with the fractional QH edges described by the Laughlin state~\cite{llf} with filling factor $\nu=1/m$, where $m$ is an odd integer.

In the present work, we broadly focus on two directions. First, we explore the transport properties of a four-site QH circuit, shown in Fig.~\ref{fig1}, which serves as a prototypical setup where higher-order backscattering processes constitute the relevant perturbations; increasing the number of sites does not introduce fundamentally new effects. Second, we introduce and analyze tunneling phenomena in multichannel realizations of multi-site mesoscopic QH circuits. Together, we establish multi-site QH circuits as a versatile and experimentally accessible platform for realizing and probing quantum criticality, interference-driven transport, and non-Fermi-liquid behavior in strongly correlated mesoscopic systems.

This paper is organized as follows. In Sec.~\ref{model} we present our bosonization-based model to describe the multi-site QH circuits. An exact solution of the model in the absence of backscattering events is given in Sec.~\ref{scatt}. Section~\ref{ele} is devoted to the construction of an effective low-energy Hamiltonian that takes into account the different tunneling processes in the considered system. Quantum critical properties of a multi-site QH circuit are discussed in Sec.~\ref{qcp}, and their transport properties in different geometries are presented in sections.~\ref{chh} and ~\ref{trr1}.  In sections~\ref{mck} and ~\ref{trr2}, we introduce the multichannel, multi-site mesoscopic QH circuits and analyze their quantum critical behaviors. Discussion of the Joule heating effects in the explored devices is given in Sec.~\ref{hee}. Finally, we conclude in Sec.~\ref{concls}. Minor mathematical details of our calculations are given in the Appendices.
\section{Model}\label{model}

The schematic of the experimental setup, consisting of four hybrid metal-semiconductor islands, is shown in Fig.~\ref{fig1}. Each island hosts a macroscopically large number of charge states, which can be tuned by applying voltages to the corresponding nearby gates (green metallic plates in the figure). The islands are connected to each other, as well as to the metallic leads by fully tunable quantum point contacts (QPCs) represented by the red dots in the figure. The free electrons in the QPCs can be modeled as pairs of counter-propagating QH edge states that are partially covered by the metallic islands. Electron propagation through these chiral edge states can conveniently be studied using bosonization of fermion operators~\cite{vonDelft1998}. In this representation, the spinless electrons in the five
QPCs are described by the quadratic Hamiltonian~\cite{Wen_1990}
\begin{align}
H_0=\frac{v_{\rm F}}{4\pi}\sum_{\alpha=1}^5\int^\infty_{-\infty}\!\!\! dx\Big[\left(\partial_x\Phi_{\alpha, {\rm R}}\right)^2+\left(\partial_x\Phi_{\alpha, {\rm L}}\right)^2\Big],\label{rud1}
\end{align}
where $\alpha$ labels the QPCs and $v_{\rm F}$ denotes the Fermi velocity. $\Phi_{\alpha, {\rm L/R}}(x, t)$ represents the bosonic field corresponding to the left/right moving chiral fermions, and satisfies the commutation relation $\left[\Phi_{\rm \alpha, R/L}(t), \Phi_{\rm \alpha, R/L}(t_1)\right]=\pm i \pi {\rm sign}(t-t_1)$. Throughout this paper we set $e=\hbar=k_{\rm B}=1$.

The Coulomb interactions in the system of four islands, each with same charging energy $E_C$, are described by the constant-interaction model~\cite{Furusaki1995b}
\begin{align}
H_C=E_C\sum_{j=1}^4\left[\hat{N}_j-N_{ j, \rm g}\right]^2,\label{rud2}
\end{align}
where $j$ labels the islands, and $N_{j,\rm  g}$ denotes a dimensionless parameter proportional to the gate voltage $V_{j, {\rm g}}$, i.e., $e N_{j,{\rm g}} = C_{j,{\rm g}} V_{j,{\rm g}}$, where $C_{j, {\rm g}}$ is the capacitance between the $j$th island and the corresponding gate. In the following, we will use the term ``dimensionless gate voltages" or simply ``gate voltages" to refer to $N_{j,{\rm g}}$. The electron number operator of the $j$th island is given, in terms of the bosonic fields evaluated at $x=0$, by
\begin{align}
\hat{N}_j(t)=\frac{1}{2\pi}\left[\Phi_{ j,\rm L}-\Phi_{ j, \rm R}+\Phi_{j+1,\rm R}-\Phi_{\rm j+1,\rm L}\right],
\end{align}
where we used the abbreviation $\Phi_{ j,\rm L/R}(x=0, t)\equiv \Phi_{ j,\rm L/R}$.
We note that in setup~\ref{fig1}, there is indeed an appreciable amount of inter-island Coulomb interaction. The cross capacitance that facilitates the charge-charge coupling between neighboring islands, however, does not allow the dc current to flow. Therefore, as long as we are interested in low-energy charge transport and remain far from a possible phase-transition, such charge-charge couplings can be safely neglected~\cite{dpk1}.

The interaction Hamiltonian~\eqref{rud2} is quadratic in the bosonic fields and can therefore be solved exactly. Moreover, if the system is fully described by equations~\eqref{rud1} and~\eqref{rud2}, the gate voltages can be gauged out, resulting in a system without charge granularity. Small but finite backscattering events mediated by the QPCs fundamentally alter this picture. Such events restore charge granularity and introduce strongly nonlinear boundary interactions of the form $H_{\rm b}=\sum_\alpha\sum_n\mathcal{V}(2n k_F) \cos \left[n\left(\Phi_{\alpha, \rm R}-\Phi_{\alpha,\rm L}\right)\right]$. For the low-energy description of a setup comprising $N$ islands, the QPC index $\alpha$ runs from 1 to  $N+1$, and the integer $n$ takes the values $n=1, 2,\cdots\lfloor \sqrt{N+1}\rfloor$. Therefore, the backscattering in the QPCs of device~\eqref{fig1}, i.e., $N=4$, is described by
\begin{align}
H_{\rm b}(t) &=\sum_{\alpha=1}^5\Big[U_\alpha\cos \left[\Phi_{\alpha, \rm R}(0, t)-\Phi_{\alpha,\rm L}(0, t)\right]\nonumber\\
&\qquad+V_\alpha\cos 2\left[\Phi_{\alpha, \rm R}(0, t)-\Phi_{\alpha,\rm L}(0, t)\right]\Big],\label{rud3}
\end{align}
where $U_\alpha\propto \mathcal{V}(2k_F)$, $V_\alpha\propto \mathcal{V}(4k_F)$ and they depend linearly and quadratically respectively, on the band cutoff $D$. In the following, we solve the interaction Hamiltonian~\eqref{rud2} exactly, and treat the nonlinearities~\eqref{rud3} perturbatively in small backscattering amplitudes~\cite{Furusaki1995b}.

\section{Exact account of the interactions}\label{scatt}
In the absence of backscattering events, the continuous flow of chiral fermions along the QH edges, described by the Hamiltonian $H_{\rm p}=H_0+H_C$, represents the plasmonic excitations propagating along the chiral edges. These excitations can be accounted for using the standard equation-of-motion approach~\cite{Slobodeniuk_2013,Sukhorukov_2016,morel2021}. This allows one to express all bosonic fields $\Phi_{\alpha,\rm R/L}(x, \omega)$ in terms of the corresponding free fields $\Phi^0_{\alpha,\rm R/L}(x, \omega)$, which satisfy the well-known commutation relations and possess standard correlators
\begin{align}
\left<\!\Phi^{(0)}_{\alpha, {\rm R/L}}(\omega)\Phi^{(0)}_{\beta, {\rm R/L}}(\omega')\!\right>_{\!\!0}\!{=}\frac{4\pi^2\delta_{\alpha\beta}}{\omega'}n\!\left(\!\!\frac{\omega'}{T}\!\right)\!\!\delta\!\left(\omega+\omega'\right)\!,\label{rud4}
\end{align}
with $n(y)=\left(e^y-1\right)^{-1}$ being Bose distribution function defined at the equilibrium temperature $T$. 

To study transport properties, it is more convenient to introduce the compact bosonic fields $\Phi_\alpha\equiv \Phi_{\alpha,\rm R}-\Phi_{\alpha,\rm L}$, along with the corresponding free fields $\Phi^0_\alpha$. It is straightforward to show, by solving the equation of motion for the Hamiltonian $H'\equiv H_0+H_C$, that the matrix $\Phi=\left(\Phi_1, \Phi_2, \Phi_3, \Phi_4, \Phi_5\right)^\top$ satisfies
\begin{align}
\Phi(\omega)=\mathbb{A}(\omega)\Phi^0(\omega)+\mathbb{N}(\omega),\label{rud4a}
\end{align}
for some matrix $\mathbb{A}$ that depends on the interaction strength and number of edge channels. The matrix $\mathbb{N}$ accounting for the gate voltages takes the form
\begin{align}
\mathbb{N}(t)=\frac{2\pi}{5}  \left(
\begin{array}{cccc}
 -4 &-3& -2 &-1 \\
 1&-3 &-2 &-1 \\
 1&2 &-2 &-1 \\
 1&2 &3 &-1 \\
 1&2 &3 &4  \\
\end{array}
\right)\left(
\begin{array}{cccc}
 N_{1,\rm g} \\
 N_{2,\rm g} \\
 N_{3,\rm g} \\
N_{4,\rm g} \\
 N_{5,\rm g}  \\
\end{array}
\right).\label{rud5}
\end{align}

To express the equation~\eqref{rud4a} into a more meaningful form, we introduce an orthogonal transformation that affects the free fields $\Phi_\alpha^0$ such that 
\begin{align}
\left(\Phi^0_1, \Phi^0_2, \Phi^0_3, \Phi^0_4, \Phi^0_5\right)^\top=\mathbb{O}\left(\Phi_a, \Phi_b, \Phi_c, \Phi_d, \Phi_e\right)^\top,\label{rud5a}
\end{align}
with an orthogonal matrix
\begin{align}
\mathbb{O}=\left(
\begin{array}{ccccc}
 \frac{1}{\sqrt{5}} & \frac{1}{\sqrt{2}} & \frac{1}{\sqrt{6}} & 0 & -\sqrt{\frac{2}{15}} \\
 \frac{1}{\sqrt{5}} & -\frac{1}{\sqrt{2}} & \frac{1}{\sqrt{6}} & 0 & -\sqrt{\frac{2}{15}} \\
 \frac{1}{\sqrt{5}} & 0 & -\sqrt{\frac{2}{3}} & 0 & -\sqrt{\frac{2}{15}} \\
 \frac{1}{\sqrt{5}} & 0 & 0 & \frac{1}{\sqrt{2}} & \sqrt{\frac{3}{10}} \\
 \frac{1}{\sqrt{5}} & 0 & 0 & -\frac{1}{\sqrt{2}} & \sqrt{\frac{3}{10}} \\
\end{array}
\right).\label{rud6}
\end{align}
Using equations~\eqref{rud4a} and~\eqref{rud5a}, we express the original bosonic fields $\Phi_\alpha$ in terms of the new free fields $\Phi_l$, $l=a, b, c, d, e$, as
\begin{align}
\Phi(\omega)=\mathbb{M}(\omega)\left(\Phi_a, \Phi_b, \Phi_c, \Phi_d, \Phi_e\right)^\top+\mathbb{N},\label{rud7}
\end{align}
The full expression of the matrix $\mathbb{M}(\omega)$ is provided in the appendix~\ref{a1}, and in the limit $\omega\tau\to 0$, where $\tau\equiv\pi/E_C$ is the usual Heisenberg time constant, it takes the form
\begin{align}
\lim_{\omega\tau\to 0}\mathbb{M}_{\alpha\beta}(\omega)=\frac{\delta_{1\beta}}{\sqrt5}.\label{rud8}
\end{align}
We note that in Eq.~\eqref{rud8}, and in the following discussion, we restrict ourselves to the regime $T\ll E_c$, where the physics is dominated by quantum fluctuations. In typical experiments, such as those in Refs.~\cite{Pierre_2015, ccd}, the base temperature is 
$T = (10 - 20) \, \rm{mK}$, and the charging energy is $E_c = (25 - 30) \, \mu\rm{eV} \sim (290 - 348) \, \rm{mK}$, which is well within the regime of our interest. The opposite regime, $T\gg E_c$,  can be investigated in the spirit of Ref.~\cite{dpk1}. From Eqs.~\eqref{rud7} and~\eqref{rud8}, it follows that out of five chiral bosonic modes, only one mode, $\Phi_a$, remains gapless, while the remaining four acquire gap due to the large charging energy. For the low-energy description, the gapped modes can be integrated out, leaving the dynamics governed solely by the free mode $\Phi_a$. Equation~\eqref{rud8} further implies that the scaling dimensions of the boundary perturbations appearing in the first and second terms of the equation~\eqref{rud3} are $1/5$ and $4/5$ respectively, and therefore relevant perturbations in RG sense.

\section{Effective low-energy Hamiltonian}\label{ele}
By integrating out the gapped modes in the equation~\eqref{rud7}, we express the backscattering Hamiltonian~\eqref{rud3} in terms of a single gapless mode $\Phi_a$. This procedure is elaborated in appendix~\ref{ca}, and the final result for the effective Hamiltonian of the four-site QH circuit writes
\begin{align}
H_{\rm eff} =&\frac{v_{\rm F}}{4\pi}\int^\infty_{-\infty}\!\!\! dx\Big[\left(\partial_x\Phi_{a, {\rm R}}\right)^2+\left(\partial_x\Phi_{a, {\rm L}}\right)^2\Big]\nonumber\\
&+|r_u|D^{1/5}\cos\left(\frac{\Phi_a}{\sqrt5}+\theta_u\right)\nonumber\\
&+|r_v|D^{4/5}\cos\left(\frac{2}{\sqrt5}\Phi_a+\theta_v\right),\label{rud9}
\end{align}
where $\theta_{u/v}$ are unimportant phase factors. In the case of common gate voltage, i.e. $N_{j, \rm g}=\ngg$, the effective reflection coefficients $|r_{u/v}|$ acquire the form

\begin{align}
r_u &=\left(\frac{5^{3/16} e^\gamma}{\tau}\right)^{4/5}\!\!\Big[\mathcal{C}_1 \mathsf{U}_1 e^{-4\pi i\ngg}+\mathcal{C}_2 \mathsf{U}_2 e^{-2\pi i\ngg}\nonumber\\
&\qquad+\mathcal{C}_3 \mathsf{U}_3 +\mathcal{C}_2 \mathsf{U}_4 e^{2\pi i\ngg}+\mathcal{C}_1 \mathsf{U}_5 e^{4\pi i\ngg}\Big],\nonumber\\
r_v &=\left(\frac{5^{3/16} e^\gamma}{\tau}\right)^{16/5}\Big[\mathcal{D}_1 \mathsf{V}_1 e^{-8\pi i\ngg}+\mathcal{D}_2 \mathsf{V}_2 e^{-4\pi i\ngg}\nonumber\\
&\qquad+\mathcal{D}_3 \mathsf{V}_3 +\mathcal{D}_2 \mathsf{V}_4 e^{4\pi i\ngg}+\mathcal{D}_1 \mathsf{V}_5 e^{8\pi i\ngg}\Big],\label{rud10}
\end{align}
where $\mathcal{C}_j$ and $\mathcal{D}_j$ are numerical coefficients arising from integrating out the gapped bosonic modes, and also include proportionality factors associated with the renormalized backscattering strengths $\mathsf{U}_j\propto U_j$ and $\mathsf{V}_j\propto V_j$; see appendix~\ref{ca} for details. The symbol $\gamma$ stands for the Euler's constant. In the following section, we analyze the quantum-critical properties of the effective Hamiltonian~\eqref{rud9}.

\section{Quantum critical behaviors}\label{qcp}
At the quantum critical point (a point in the parameter space of the gate voltage and barrier transparencies), all backscattering processes interfere destructively to nullify the backscattering terms in the Hamiltonian~\eqref{rud9}. At this point, the unitary conductance is achieved (see below). We note that since both backscattering terms are relevant, one is required to find proper parameters to nullify both $|r_u|$ and $|r_v|$. Although there are several parameters in the equation~\eqref{rud9} that can be tunned experimentally, we focus below on a few representative cases.

First consider the situation where only the leftmost and rightmost QPCs, i.e., the first and fifth, are in quasi-ballistics regime with equal reflection amplitudes $\mathsf{U}_1=\mathsf{U}_5$ and $\mathsf{V}_1=\mathsf{V}_5$. Since all the inner QPCs are fully open ( $\mathsf{U}_{2, 3, 4}=0=\mathsf{V}_{2, 3, 4}$), the effective reflection coefficients take the form $|r_u|\propto \cos4\pi\ngg$ and $|r_v|\propto \cos8\pi \ngg$. Due to the different gate-voltage periodicities of these coefficients, they never vanish simultaneously for any gate voltage. This special realization of device~\ref{fig1} thus does not support a zero-temperature quantum critical point, i.e., only a perturbative regime exists, and the dominant scaling behaviors of transport and thermodynamic quantities are determined by the relative strengths of two corrections in equation~\eqref{rud9}.

We now discuss the left-right symmetric realization of device~\ref{fig1}, in which the barrier transparencies are chosen such that $\mathcal{C}_1\mathsf{U}_1=\mathcal{C}_2\mathsf{U}_2=\mathcal{C}_4\mathsf{U}_4
=\mathcal{C}_5\mathsf{U}_5$ and $\mathcal{C}_3\mathsf{U}_3/\mathcal{C}_1\mathsf{U}_1=\delta_u$, and similarly $\mathcal{D}_1\mathsf{V}_1=\mathcal{D}_2\mathsf{V}_2=\mathcal{D}_4\mathsf{V}_4
=\mathcal{D}_5\mathsf{V}_5$ and $\mathcal{D}_3\mathsf{V}_3/\mathcal{D}_1\mathsf{V}_1=\delta_v$. In this case, we have $|r_u|\propto \left[\delta_u+2\cos2\pi\ngg+2\cos 4\pi\ngg\right]$ and $|r_v|\propto \left[\delta_v+2\cos4\pi\ngg+2\cos 8\pi\ngg\right]$. It is therefore possible to find the parameter space $\left(\delta_u, \delta_v, \ngg\right)$ where $|r_u|$ and $|r_v|$ vanish simultaneously. As an example, for a given value of $\delta_u$, the fine-tuned gate voltage
\begin{align}
\ngg^*=\frac{\tan ^{-1}\left[\frac{1}{4} \left(\sqrt{9{-}4 \delta_u }{-}1\right),\frac{\sqrt{2 \delta_u +\sqrt{9-4 \delta_u }+3}}{2 \sqrt{2}}\right]}{2 \pi },\label{rud11}
\end{align}
and $\delta_v$ chosen such that
\begin{align}
\delta^*_v=3 \delta_u +\sqrt{9-4 \delta_u}-1-\delta_u ^2-\delta_u\sqrt{9-4 \delta_u },
\end{align}
together constitute one of the quantum critical points. As seen from equation~\eqref{rud11}, the critical regime emerges only for $0<\delta_u\leq 9/4$, and for larger $\delta_u$ different critical points merge and annihilate each other. One can numerically solve the zero condition of the reflection coefficients given in equation~\eqref{rud10} to obtain the full structure of the critical points for general barrier transparencies and gate voltages. In the following, we show that these features can be accessed via transport measurements.
\begin{figure*}[ht!]
\begin{center}
\includegraphics[scale=0.35]{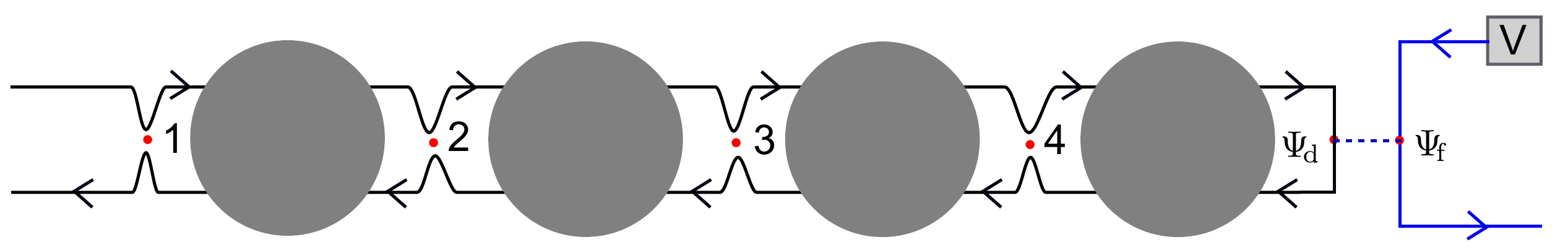}
\caption{Schematic representation of the setup~\ref{fig1} in the asymmetric tunneling regime (see text for details).}\label{fig2}
\end{center}
\end{figure*}

\section{Charge current}\label{chh}
Here we evaluate the charge current measured at the rightmost QPC (QPC 5), described by the operator $\hat{I}=-\frac{1}{2\pi}\frac{\partial\Phi_5}{\partial t}$. As discussed earlier, all fields are function of $\Phi_a$, and since $\Phi_a$ is the only gapless mode, the application of the voltage bias $V\ll E_C$ amounts to the dressing such that $\Phi_a\to \Phi_a-\left(Vt/\sqrt5\right)$. Therefore, in the fully ballistic limit, the charge current is given by
\begin{align}
I=-\frac{1}{2\pi}\left<\frac{\partial \Phi_5}{\partial t}\right>=\frac{1}{5}\frac{V}{2\pi},\label{rud12}
\end{align}
which is equivalent to the conductance of five quantum resistors (with resistors $R_q=2\pi$) connected in series.

The backscattering correction to the charge current is accounted for by the usual second-order perturbation theory

\begin{align}
\delta I &={-}\int^t_{-\infty} \!\!\!\!dt_1\!\! \int^{t_1}_{-\infty}\!\!\! dt_2\!\left<\!\Big[H_{\rm eff}(t_2),\! \left[H_{\rm eff}(t_1), \hat{I}(t)\right]\!\Big]\!\right>.\label{rud13}
\end{align}
Since the two backscattering sectors in equation~\eqref{rud9} do no produce finite inter-sector interfere contribution at the lowest order (due to the imbalance of charge neutrality condition), equation~\eqref{rud13} simplifies to
\begin{align}
\delta I &=\frac{1}{5}\int^\infty_{-\infty} dt\left<\left[\mathcal{A}_u(t), \mathcal{A}^\dagger_u(0)\right]\right>\nonumber\\
&\;\;\;+\frac{2}{5}\int^\infty_{-\infty} dt\left<\left[\mathcal{A}_v(t), \mathcal{A}^\dagger_v(0)\right]\right>,\label{rud14}
\end{align}
where the operators $\mathcal{A}_{u/v}$ are defined as
\begin{align}
\mathcal{A}_u=\frac{|r_u|}{2}e^{\frac{i\Phi_a}{\sqrt5}}, \;\;\mathcal{A}_v=\frac{|r_v|}{2}e^{\frac{2i\Phi_a}{\sqrt5}}.
\end{align}
Owing to have only a free field $\Phi_a$, it is straightforward to evaluate the averages in equation~\eqref{rud14}. The resulting total charge current writes
\begin{align}
I=\frac{1}{5}\frac{V}{2\pi}\Bigg(1{-}\left[\frac{\pi^2 5^{3/5}|r_u|^2 }{V^{8/5} \Gamma \left(\frac{2}{5}\right)}{+}\frac{2\pi^2\left(\frac{2}{5}\right)^{3/5}|r_v|^2 }{V^{2/5} \Gamma \left(\frac{8}{5}\right)}\right]\Bigg),\label{rud15}
\end{align}
where $\Gamma(y)$ denote the Gamma function with argument $y$. Precisely at the critical point discussed above, the correction given by the second term in equation~\eqref{rud15} vanishes, and the perturbative treatment remains valid down to zero energies. We note that the linear response $V\to 0$ correction to the conductance can be obtained, up to a trivial numerical prefactor, by replacing $V\to T$ in the second part of the equation~\eqref{rud15}. The backscattering correction in equation~\eqref{rud15} can also be expressed as $\sim \left(T^*_u/V\right)^{8/5}+\left(T^*_u/V\right)^{2/5}$, where we introduced the crossover scales $T^*_u\propto |r_u|^{5/4}$ and $T^*_v\propto |r_v|^{5}$. In the general case, the validity of equation~\eqref{rud15} is restricted to the region $T^*_{u, v}\ll (V, T)$. At the quantum critical point, both crossover scales vanish, and a small detuning of the gate voltage from the critical value $N^*_{\rm g}$~\eqref{rud11} gives the dominant scaling behavior $T^*_u\sim \delta N_{\rm g}^{5/4}$, provided that both $|r_u|$ and $|r_v|$ remain finite in the vicinity of critical point. Here, $\delta N_{\rm g}=N_{\rm g}-N^*_{\rm g}$. If, for a given detuning, $T^*_u=0$, the scaling is much weaker, with $T^*_v\sim \delta N_{\rm g}^{5}$.

\section{Transport properties in the tunneling regime}\label{trr1}
We now consider the case where the rightmost QPC in Fig.~\ref{fig1} operates in the tunneling regime, while the remaining QPCs are fully ballistic or in the quasi-ballistic regime as schematically shown in Fig.~\ref{fig2}. We assume that the contact conductance, i.e. $G_c=|\gamma_t|^2/2\pi$, where $|\gamma_t|$ is the tunneling amplitude of the right contact, is the smallest energy scale in the system. In this case, the tunneling current can be expressed in terms of the electron Greens function $G^>(t)=-i\Big<\Psi_d(t)\Psi_d^\dagger(0)\Big>$ of the fourth grain defined at the tunneling position $x=0$, and that of the usual free electron Greens functions~\cite{amm}. For the grain, the explicit solution of the equation of motion gives the greater Greens function in the form
\begin{align}
G^>(t)&=g^>(t)\exp\Bigg[\int^\infty_{-\infty}\frac{d\omega}{\omega}Z(\omega)\frac{e^{-i\omega t}-1}{1-e^{-\omega/T}}\Bigg],
\end{align}
where $g^>(t)=-i\Big<\Psi_f(t)\Psi_f^\dagger(0)\Big>$ is the  greater Greens function describing the free electrons, and $Z(\omega)$ is the usual frequency-dependent environmental function given by~\cite{sochin, Averin1992}
\begin{align}
Z(\omega)=\frac{8+68 \tau ^2 \omega ^2+28 \tau ^4 \omega ^4+2 \tau ^6 \omega ^6}{1+70 \tau ^2 \omega ^2+87 \tau ^4 \omega ^4+19 \tau ^6 \omega ^6+\tau ^8 \omega ^8}.
\end{align}
When a voltage $V$ is applied to the free edge channel on the right, the tunneling current can be cast into the form~\footnote{The charge current takes the standard form $I=|\gamma_t|^2\int^\infty_{-\infty} dt\;e^{iV t}\Big[g^<(t)G^<(t)-g^>(t)G^>(t)\Big]$.}
\begin{align}
I(V, T) &=-G_C\pi T^2\sinh\left(\frac{V}{2T}\right)\nonumber\\
&\qquad\times\int^\infty_{-\infty} \frac{dt}{\cosh^2\pi T t}\exp\left[\mathcal{J}(t)+i V t\right],\label{rud17}
\end{align}
where, we introduced a new function $\mathcal{J}(t)$ defined by
\begin{align}
\mathcal{J}(t)=\int^\infty_{0} \frac{d\omega}{\omega} Z(\omega)\frac{\cos\omega t-\cosh(\omega/2T)}{\sinh(\omega/2T)}.
\end{align}
In the asymptotic limit of $T\ll E_C$, the equation~\eqref{rud17} provides weakly temperature-dependent linear-response conductance $G(T)=\mathcal{F} G_c(T/E_C)^8$, where $\mathcal{F}$ is a constant of order unity. The temperature scaling exponent is governed by four pairs of electron-hole excitations each contributing $T^2$ to the conductance, and thus the setup with large number of sites becomes essentially an electronic insulator at low temperature. All the previous treatments can be applied to compute the finite backscattering corrections to equation~\eqref{rud17}. Inclusion of such backscattering corrections will amount to the addition of linear in $|r_{u/v}|$ terms to the constant $\mathcal{F}$.
\begin{figure*}[ht!]
\begin{center}
\includegraphics[scale=0.38]{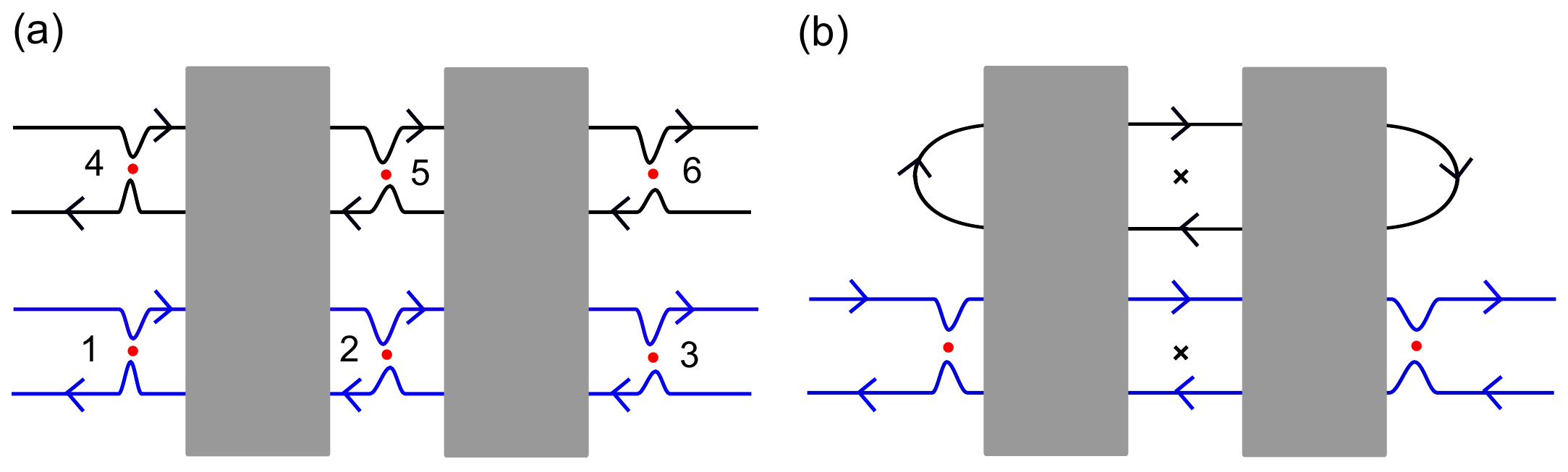}
\caption{(a) Typical realization of circuits with two channels and two sites in a quantum Hall system. (b) Setup (a) with one of the channels looped. The red dot represents finite backscattering, while ithe cross symbol indicates that the corresponding QPC is fully opened. Note that the gates are not explicitly shown (see the text for details).}\label{fig3}
\end{center}
\end{figure*}

We note that, although we assumed that the rightmost QPC operated in a tunneling regime, the above results remain identical to that for the setup with a tunneling barrier at the position of $\alpha$th QPC other than the rightmost one. If a barrier divides a circuit into two parts having impedances $Z_1$ and $Z_2$, then for all the divisions, irrespective of its position, the quantity $\lim_{\omega\to 0}(Z_1+Z_2)$ that governs the scaling behavior of the linear conductance remains the same.

We also note that if the leftmost QPC in the setup~\ref{fig2} is also operated in the tunneling regime, one needs to consider the charge quantization effect even if all the middle QPCs are fully ballistic. In this case, the charge in each grain is not separately quantized, but the charge in the big composite grain formed by four islands gets quantized, and interesting Kondo features emerge~\cite{addition1}.

\section{Multi-channel mesoscopic quantum Hall circuits}\label{mck}

In the previous sections, we studied the QH circuits in which each grain accommodates a single channel to the left and also to the right of it. More generally, it is possible to engineer a mesoscopic QH circuit with multiple edge channels traversing each metal grain~\cite{dde1}. The fully ballistic regime of such multichannel QH circuits can be understood as follows. A portion of QH circuit outside the metal grain that consists of left and right moving edges can be described by a chiral bosonic field. Therefore, for the setup with $N$ identical grains, each hosting $M$ edge channels, there are in total $M(N+1)$ bosonic modes describing the corresponding plasmonic excitations. At energies much lower than the charging energy, and provided that the inter-grain Coulomb interactions are negligible, $N$ of the $M(N+1)$ bosonic modes are gapped by the charging effects and can be integrated out. The corresponding low-energy effective Hamiltonian is thus expressed in terms of the remaining $N_{\rm f}=\left[M(N+1)-N\right]$ gapless modes. Standard circuit analysis then shows that the conductance of each free channel takes the value $G_0=M/[2\pi(N+1)]$.

When $N_{\rm f}>1$, the problem can in principle be mapped onto  that of a particle moving in a multidimensional periodic potential subject to dissipation~\cite{Yi, Yi1}. The usual perturbative calculations at large transparencies are generally unstable, in the sense that as temperature/voltage tends to zero, the perturbative corrections blow up. To explicitly demonstrate this behavior, we now study a typical multichannel QH circuit as schematically shown in Fig.~\ref{fig3}(a). It consists of two islands and a total of six edge channels. For the clarity of presentation, we refer to the portion of the circuit to the left of the first island as the left part, that between the two grains as the middle part and that right of the second grain as the right part. Each part accommodates two channels (blue and black), here for generality, implemented via six QPCs as shown in Fig.~\ref{fig3}(a).

We define six compact bosonic fields $\Phi_\alpha=\Phi_{\alpha,\rm R}-\Phi_{\alpha, \rm L},\;\alpha=1, 2, \cdots, 6$, that fully represent the entire portion of the circuit not covered by the grains. A constant interaction model, similar to that detailed in the previous sections, can easily be constructed. The boundary perturbations, describing weak backscattering events in six interspaced QPCs, can be written in the standard form
\begin{align}
\mathcal{H}_{\rm b}=\sum_{\alpha=1}^6\frac{D|r_\alpha|}{\pi}\cos\Phi_\alpha\left(0, t\right),\label{rud18}
\end{align}
where $|r_\alpha|$ represents the weak backscattering amplitude of the $\alpha$th QPC.

To study the edge magnetoplasmon dynamics in the device~\ref{fig3}(a), it is more convenient to establish charge and spin modes in the left, middle and right parts of the circuit. To this end, we define the new fields $\Phi_{\rm LC/LS}=\left(\Phi_1\pm\Phi_4\right)/\sqrt2$, $\Phi_{\rm MC/MS}=\left(\Phi_2\pm\Phi_5\right)/\sqrt2$ and $\Phi_{\rm RC/RS}=\left(\Phi_3\pm\Phi_6\right)/\sqrt2$. We now focus to the symmetric device such that $|r_1|=|r_4|=|r_{\rm L}|$, $|r_2|=|r_5|=|r_{\rm M}|$, $ |r_3|=|r_6|=|r_{\rm R}|$. In this case, the backscattering Hamiltonian~\eqref{rud18} can be written as
\begin{align}
\mathcal{H}_{\rm b}=\sum_{j=\rm L, M, R}\frac{2D|r_j|}{\pi}\cos\left(\frac{\Phi_{j\rm C}}{\sqrt2}\right)\cos\left(\frac{\Phi_{j\rm S}}{\sqrt2}\right),\label{rud20}
\end{align}
which shows the spin-charge separation. While the spin modes $\Phi_{j\rm S}$ are not affected by charging energy and remain essentially free, the modes $\Phi_{j\rm C}$ acquire charge dynamics. Indeed, among the three charge modes, $\Phi_{j\rm C}$, only their symmetrical combination remains free and the two other orthogonal modes are gapped by large charging energy. To show this, we followed the usual equation of motion approach (as illustrated in great detail in previous sections), envisioned a number of orthogonal transformations and finally expressed the modes $\Phi_{j\rm C}(\omega)$ in terms of three incoming fields $\Phi_{\rm f, g, h}(\omega)$ and two gate voltages $N_{1/1,\rm g}$ such that

\begin{align}
\left(
\begin{array}{c}
 \Phi_{\rm LC} \\
 \Phi_{\rm MC}\\
 \Phi_{\rm RC} \\
\end{array}
\right) &=\left(
\begin{array}{ccc}
 \frac{\sqrt{6} (1-4 \mathcal{K}_1)}{12 \mathcal{K}_1+6} & \frac{1}{\sqrt{3}} & \frac{1-4 \mathcal{K}_1}{\sqrt{2} (2 \mathcal{K}-1)} \\
 \frac{\sqrt{6} (4 \mathcal{K}_1-1)}{6 \mathcal{K}_1+3} & \frac{1}{\sqrt{3}} & 0 \\
 \frac{\sqrt{6} (1-4 \mathcal{K}_1)}{12 \mathcal{K}_1+6} & \frac{1}{\sqrt{3}} & \frac{4 \mathcal{K}_1-1}{\sqrt{2} (2 \mathcal{K}_1-1)} \\
\end{array}
\right)\left(
\begin{array}{c}
 \Phi_{\rm f} \\
 \Phi_{\rm g }\\
 \Phi_{\rm h} \\
\end{array}
\right)\nonumber\\
&\qquad+2 \pi  \left(
\begin{array}{c}
 \frac{\sqrt{2} \mathcal{K} (2 \mathcal{K}_1 N_{2,\rm g}+N_{1,\rm g})}{4 \mathcal{K}^2_1-1} \\
 \frac{\sqrt{2} \mathcal{K} (N_{1,\rm g}-N_{2,\rm g})}{2 \mathcal{K}_1+1} \\
 -\frac{\sqrt{2} \mathcal{K}_1 (2 \mathcal{K}_1 N_{1,\rm g}+N_{2,\rm g})}{4 \mathcal{K}^2_1-1} \\
\end{array}
\right),\label{rud21}
\end{align}
with $1/\mathcal{K}_1=4-i\omega\tau$. From the equation~\eqref{rud21}, it is clear that at the limit $\omega\tau\to 0$, all the charge modes $\Phi_{j\rm C}$ are in turn a function of only a global charge mode $\Phi_{\rm g}$, and thus the other two fields $\Phi_{\rm f, h}$ can be integrated out. At low energies, the equation~\eqref{rud20} then takes the effective form 
\begin{align}
\mathcal{H}_{\rm eff}=\frac{2 D}{\pi}\frac{\sqrt[3]{6}}{\sqrt[4]{3}} \left(\frac{e^\gamma}{D \tau}\right)^{1/3}  \mathscr{H},\label{rud22}
\end{align}
where the operator $\mathscr{H}$ is defined by
\begin{align}
\mathscr{H} &=|r_M|3^{1/4}\cos\left[\frac{\Phi_{\rm g}}{\sqrt6}{+}\frac{\pi(N_{1,\rm g}{-}N_{2,\rm g})}{3}\right]\cos\left(\frac{\Phi_{\rm MS}}{\sqrt2}\right)\nonumber\\
&+|r_L|\cos\left[\frac{\Phi_{\rm g}}{\sqrt6}{-}\frac{\pi(2N_{1\rm g}{+}N_{2,\rm g})}{3}\right]\cos\left(\frac{\Phi_{\rm LS}}{\sqrt2}\right)\nonumber\\
&+|r_R|\cos\left[\frac{\Phi_{\rm g}}{\sqrt6}{+}\frac{\pi(2N_{1\rm g}{+}N_{2,\rm g})}{3}\right]\cos\left(\frac{\Phi_{\rm LS}}{\sqrt2}\right).\label{rud23}
\end{align}
We note that all three terms in the equation~\eqref{rud22} have scaling dimensions $2/3$. The four gapless modes, $\Phi_{\rm g}$ and $\Phi_{j\rm S}$ consistently need to be taken into account to describe the dynamics of the device~\ref{fig3}(a). For the device with $N$ grains each having $M$ channels (to the left or right), one can repeat the above procedure to arrive at the effective Hamiltonian with $M(N+1)-N$ free bosonic modes.

Having arrived at equation~\eqref{rud23}, it is straightforward to calculate its transport and thermodynamic properties at large transparencies. As an illustration, the conductance carried by the free charge mode $\Phi_{\rm g}$ takes the form
\begin{equation}
G=\frac{2}{3}\frac{1}{2\pi}\Bigg[1-\frac{2^{5/3} 3^{1/6}}{\pi^{1/6}}\left(\frac{e^\gamma E_C}{\pi T}\right)^{2/3}\frac{\Gamma(2/3)}{\Gamma(1/6)}\mathcal{X}\Bigg],\label{rud24}
\end{equation}
where we introduced the function $\mathcal{X}$ defined by
\begin{align}
\mathcal{X}=|r_L|^2+\sqrt3 |r_M|^2+|r_R|^2.
\end{align}
The conductance~\eqref{rud24} is always smaller than the corresponding unitary value, even at the resonance. As $T\to 0$, the correction to the conductance diverges, signaling the breaking of perturbative treatments. The corresponding non-perturbative solutions could be obtained to some extent by using the quantum Brownian motion approach~\cite{Yi, Yi1} or the functional renormalization group method~\cite{moranew}.

\section{Exotic quantum critical points in multi-channel QH circuits}\label{trr2}
As long as more than one gapless mode governs the low-energy dynamics, a mapping to the boundary sine-Gordon model is no longer applicable. Nonetheless, following Ref.~\cite{dde1}, we show that in the multichannel QH circuits the number of gapless modes can be reduced to one by looping a selected number of edge channels. This method enables the realization of zero-temperature quantum critical phenomena characterized by a general scaling dimension $\eta<1/2$, where $\eta$ depends on both the number of grains and the number of looped edge channels.

To illustrate the looping method, we consider a simples setup as shown in Fig.~\ref{fig3}(b), obtained by looping the second edge channel of Fig.~\ref{fig3}(a) back onto the grains~\cite{looping}. The key idea is that transport through the left and right QPCs (QPC 1 and 3 respectively) is unaffected by the presence of the looped part (on the left and right) of the circuit. Therefore, if both channels in the middle part are fully ballistic, they merely provide impedances to the incoming electrons at the left and right QPCs,and
the entire setup becomes effectively equivalent to a single-grain device with two QPCs. The crucial difference here is that the number of connecting ballistic channels contributes an additional impedance, which in turn controls the scaling dimension of the boundary electron operators.

Following the equation-of-motion approach to account for the charging energy part of the Hamiltonian, one can express all bosonic fields in terms of incoming fields. The fields $\Phi_{\rm o}$ describing the two looped parts of the circuit, do not directly enter the transport problem through QPCs 1 and 3, and possess the correlator of the form
\begin{align}
\left<e^{i\Phi_{\rm o}(t)}e^{-i\Phi_{\rm o}(0)}\right>=\exp\int^\infty_{-\infty}\!\!\!\frac{d\omega}{\omega}\frac{1+\tilde{Z}(\omega)}{1{-}e^{-\omega/T}}\left(e^{-i\omega t}{-}1\right),\nonumber
\end{align}
with $\tilde{Z}$ defined as
\begin{align}
\tilde{Z}(\omega)=\frac{6 \left(\tau ^2 \omega ^2+5\right)}{\left(\tau ^2 \omega ^2+1\right) \left(\tau ^2 \omega ^2+25\right)}.
\end{align}
We describe the remaining four channels by the bosonic fields $\Phi_{1, 2, 3, 5}(\omega)$. These fields can in turn be expressed in terms of four suitably defined incoming fields $\Phi_{\rm p, q, r, s}(\omega)$ as 
\begin{align}
\left(
\begin{array}{c}
 \Phi _1 \\
 \Phi _2 \\
 \Phi _3 \\
 \Phi _5 \\
\end{array}
\right)\!\!{=}\!\!\left(
\begin{array}{cccc}
 \frac{4 \mathcal{K}_1-1}{\sqrt{10} (\mathcal{K}_1+1)} & 0 & \sqrt{\frac{2}{5}} & \frac{1-4 \mathcal{K}_1}{\sqrt{2} (3 \mathcal{K}_1-1)} \\
 \frac{\sqrt{\frac{2}{5}} (1-4 \mathcal{K}_1)}{\mathcal{K}_1+1} & -\frac{1}{\sqrt{2}} & \frac{1}{\sqrt{10}} & 0 \\
 \frac{4 \mathcal{K}_1-1}{\sqrt{10} (\mathcal{K}_1+1)} & 0 & \sqrt{\frac{2}{5}} & \frac{4 \mathcal{K}_1-1}{\sqrt{2} (3 \mathcal{K}_1-1)} \\
 \frac{\sqrt{\frac{2}{5}} (1-4 \mathcal{K}_1)}{\mathcal{K}_1+1} & \frac{1}{\sqrt{2}} & \frac{1}{\sqrt{10}} & 0 \\
\end{array}
\right)\!\!\!\left(
\begin{array}{c}
 \Phi _{\rm p} \\
 \Phi _{\rm q} \\
 \Phi _{\rm r} \\
 \Phi _{\rm s} \\
\end{array}
\right),\label{rud25}
\end{align}
where the gates voltages are not explicitly shown. 

From the equation~\eqref{rud25} and remembering the definition $1/\mathcal{K}_1=4-i\omega\tau$, we see that, at low energy $\omega\tau\to 0$, all the modes $\Phi_{1, 2, 3, 5}$ can be expressed in terms of a single gapless mode $\Phi_{\rm r}$. After integrating out the remaining gapped modes, we finally arrive at the effective Hamiltonian describing the low-energy physics of the looped two-site QH circuit~\cite{dde1}
\begin{align}
H'_{\rm eff}=D^{2/5}&\left(\frac{e^\gamma 5^{\frac{1}{6}}}{\pi^{\frac{5}{3}}\tau}\right)^{3/5}\mathscr{H}',\label{rud26}
\end{align}
where, we defined $\mathscr{H}'$ as
\begin{align}
\mathscr{H}'{=}|r_{\rm L}|\cos\left(\sqrt{\frac{2}{5}}\Phi_{\rm r}{-}N_{\rm L, g}\right){+}|r_{\rm R}|\cos\left(\!\sqrt{\frac{2}{5}}\Phi_{\rm r}{+}N_{\rm R, g}\!\right),\nonumber
\end{align}
with the gate voltage parameters $N_{\rm L, g}=2\pi\left(3N_{\rm 1g}+2N_{\rm 2 g}\right)/5$ and $N_{\rm R, g}=2\pi\left(2N_{\rm 1g}+3N_{\rm 2 g}\right)/5$. The boundary term~\eqref{rud26} having scaling dimensions $2/5$ supports a zero-temperature quantum critical point that can be reached by varying gate voltages and/or backscattering amplitudes. In addition, here the boundary sine-Gordon model can be exploited to uncover all transport and thermodynamic properties. In the case of $N$ grains and $M$ looped channels, see Fig.~\ref{fig3}(b) for $N=2$ and $M=1$, the backscattering in the left and right QPCs can be described by a single gapless field, and the leading-order perturbations have the scaling dimension $\nu_{\rm o}=(M+1)/(2M+N+1)$. All these different realizations support distinct zero-temperature quantum critical points, where for $\nu_{\rm o}<1/4$, the higher order backscattering terms also have to be taken into account. Therefore, channel looping in multi-site QH circuits provides a powerful and experimentally accessible route to engineer exotic quantum critical points, yielding far-reaching insights into non-Fermi liquid behavior.

\section{Heating effects in a multi-site QH circuit}\label{hee}
In section~\ref{chh}, we presented results for the transport properties of the four-site QH circuit in a fully non-equilibrium regime driven by an voltage bias. In that analysis, we neglected the effects coming from the heating of the grains by injected voltage, namely the Joule heat $J_V=V^2/2 R_q$ carried by each incoming chiral edge due to the corresponding voltage drop $V$. Likewise, at a given temperature $T$, the heat current carried by a free chiral edge described by the bosonic field $\Phi^0$ takes the form

\begin{align}
J=\frac{R_q}{2(2\pi)^4}\int^\infty_{-\infty} d\omega \omega^2\left<\Phi^0(\omega)\Phi^0(-\omega)\right>,\label{aa}
\end{align}

Using the expression of the free correlator given in equation~\eqref{rud4}, followed by subtraction of corresponding vacuum contribution, the above equation provides the heat current carried by a free chiral edge at temperature $T$ as $J_T=\pi^2 T^2/6 R_q$. Thus, because of $J_{ V}$ and $J_{T}$, in general, the grains get indeed heated. This heating is always opposed by heat dissipation to phonons in the sample. Due to the heating and subsequent cooling processes, together with the complex charging energy profile, the grains generally acquire a nonequilibrium temperature distribution, which can be determined by solving the full kinetic equation~\cite{heating1, heating}.

To obtain rough estimates for the grain temperatures in the presence of applied voltage, we here follow a rather simplifying approach. We assume that equilibrium is established within each grain at some unknown temperature $T_j, j=1, 2, 3, 4$, and neglect the back-action (cooling) produced by phonons. In this case, a set of four heat balance equations can be solved. The grain temperatures and the temperature $T$ of the incoming edges in setup~\ref{fig1} enter the correlators of the corresponding bosonic fields. Namely, the correlators $\left<\Phi_{ j+1,\rm R}^0(\omega)\Phi_{ j+1,\rm R}^0(\omega')\right>$ and $\left<\Phi_{ j,\rm L}^0(\omega)\Phi_{ j,\rm L}^0(\omega')\right>$ satisfy equation~\eqref{rud4} with $T=T_j$, and similarly that for the incoming fields $\left<\Phi_{1, \rm R}^0(\omega)\Phi_{1,\rm R}^0(\omega')\right>$ and $\left<\Phi_{5,\rm L}^0(\omega)\Phi_{5 ,\rm L}^0(\omega')\right>$ at temperature $T$.

The heat current carried by each edge in Fig.~\eqref{fig1} can be calculated by first expressing the fields $\Phi_{\rm \alpha, R/L}(\omega)$ in terms of the corresponding incoming fields $\Phi^0_{\rm \alpha, R/L}(\omega)$ and exploiting equation~\eqref{rud4} at the corresponding temperatures. However, the general expressions for the heat currents become complicated, and one needs to rely on numerically solving such heat balance equations. Nevertheless, progress can be made to have a rough estimate of temperatures in the asymptotic limit $E_C\to\infty$. In this case, the equation of motion can be solved to show a tentative relation
\begin{align}
\mathbb{P}^{\rm out}(\omega)=\mathbb{S}(\omega)\mathbb{P}^{\rm in}(\omega),\label{rud29}
\end{align}
where we introduced $\mathbb{P}^{\rm out}=\left(\Phi_{\rm 1, R}, \Phi_{\rm 1 L},\Phi_{\rm 2R},\Phi_{\rm 2L}\cdots\right)^\top$, $\mathbb{P}^{\rm in}=\mathbb{P}^{\rm out}(\Phi_{\rm \alpha, R/L}\to\Phi^0_{\rm \alpha, R/L})$ and $\mathbb{S}$ defined by
\begin{align}
\mathbb{S}=\frac{1}{5}\left(
\begin{array}{cccccccccc}
 0 & 5 & 0 & 0 & 0 & 0 & 0 & 0 & 0 & 0 \\
 1 & 4 & -1 & 1 & -1 & 1 & -1 & 1 & -1 & 1 \\
 -4 & 4 & 4 & 1 & -1 & 1 & -1 & 1 & -1 & 1 \\
 -3 & 3 & 3 & 2 & -2 & 2 & -2 & 2 & -2 & 2 \\
 -3 & 3 & 3 & -3 & 3 & 2 & -2 & 2 & -2 & 2 \\
 -2 & 2 & 2 & -2 & 2 & 3 & -3 & 3 & -3 & 3 \\
 -2 & 2 & 2 & -2 & 2 & -2 & 2 & 3 & -3 & 3 \\
 -1 & 1 & 1 & -1 & 1 & -1 & 1 & 4 & -4 & 4 \\
 -1 & 1 & 1 & -1 & 1 & -1 & 1 & -1 & 1 & 4 \\
 0 & 0 & 0 & 0 & 0 & 0 & 0 & 0 & 0 & 5 \\
\end{array}
\right).\nonumber
\end{align}

To solve the heat-balance equations, we also need to know the voltage distribution among the edge channels after biasing the field $\Phi_{\rm 1,R}$ by the voltage $V$. Such distributions can easily be obtained just by demanding that the current carried by the fields $\Phi_{\rm \alpha, R}-\Phi_{\rm \alpha, L}$ is the same for all $\alpha$. In this way, we obtain the voltage carried by the right and left moving edges as
\begin{align}
({\rm \alpha, R})\to \frac{5-(\alpha-1)}{5}V,\;\;({\rm \alpha, L})\to \frac{5-\alpha}{5}V.\label{rud30}
\end{align}
Using equations~\eqref{rud4},~\eqref{aa} and~\eqref{rud29}, we solve the heat balance equations to get the temperatures of the grains
\begin{align}
T_j^2\simeq T^2+\frac{3V^2}{2\pi^2}.\label{rud31}
\end{align}
In the case where the two incoming edge channels described by the fields $\Phi_{\rm 1, R}$ and $\Phi_{\rm 5, L}$ are kept at different temperatures $T_{s/d}$, the temperature in the above equation will be modified as $T^2\to (T^2_s+T^2_d)/2$. [The asymptotic limit $T^2_j= (T^2_s+T^2_d)/2$ is an expected result, since in the fully ballistic limit, the multi-site QH circuit can be thought of as an equivalent circuit representing a big composite grain coupled to two reservoirs. The interpretation holds true under the assumption that the edge states between grains are fully covered by the connecting grains and are responsible only for proving impedances to the incoming electrons.] Similarly, in the case that the incoming edges are biased by a different voltage, $V_{s/d}$, the voltage in the equation~\eqref{rud31} changes to $V\to V_s-V_d$. As seen from equation~\eqref{rud31}, in the out-of-equilibrium setting of setup~\ref{fig3}, the heating effects are indeed significant. In all the linear response calculations presented earlier, the inclusion of heating effects leads to vanishingly small corrections that can thus be neglected.

We note that in the above calculations, we neglected the charge-charge coupling between the neighboring grains. In the realistic situations, such simplifying assumption however fails, and indeed an appreciable inter-island Coulomb interactions, that could even be of the same order of magnitude as that of intra-island, always exist~\cite{dpk1}. The inter-island Coulomb interactions can be modeled by the cross-capacitances $C_c$. These capacitances do not allow the DC charge transport, but mediate an exchange of energy or heat between the neighboring grains. Therefore, inter-island Coulomb interactions strongly affect the different aspects of heat transport in multi-site QH circuits, and must be taken into account to obtain quantitative predictions for corresponding heat-transport properties. At the technical label, when the cross-capacitance $C_c$ and direct-capacitance $C$ are both present, the estimation of heating effects will not lead to a simple result~\eqref{rud31}, since both capacitances can not tend to zero simultaneously~\cite{dragg}. One can also imagine a theoretical scenario where the grains are really far apart such that inter-grain interactions can be neglected. In this case, one is required to consider the finite segment of the circuit between the neighboring grains. The latter problem is then equivalent to that of a short quantum wire connected to leads with finite charging energies, and is known to have different physics than the usual QH circuits.

\section{Conclusion}\label{concls}
We explored the transport properties of multi-site QH circuits, with particular focus on the prototypical four-site setup, where higher-order backscattering processes constitute the relevant perturbations. We showed that consideration of such a higher order processes is crucial for correctly identifying the quantum critical features of the corresponding model. These critical properties are shown to be directly imprinted in the charge current flowing through the circuit in response to an applied voltage bias. We also investigated tunneling phenomena in the multichannel, multi-site mesoscopic QH circuits by explicitly evaluating their transport properties and demonstrated robust non-Fermi liquid behavior. We further proposed a coveted pathway, based on looping QH edge channels, to realize exotic zero-temperature critical phenomena in multi-site QH circuits. This approach allows the realization of a wide range of critical exponents by adjusting the number of grains and that of the looped edge channels. Finally, we quantified the effects of Joule heating in the explored devices. Our work paves the way for simulating a broad range of strongly correlated phenomena in nanoengineered electronic circuit with well-defined quantum Hall channels.

\section*{Acknowledgements}
Fruitful discussions with Frederic Pierre are gratefully acknowledged. 
\appendix

\section{Edge magnetoplasmon scattering}\label{a1}
The solution of the equation of motion for the Hamiltonian $H'=H_0+H_C$, where $H_0$ and $H_C$ are given in equations~\eqref{rud1} and~\eqref{rud2} respectively, followed by the rotation~\eqref{rud6}, can be arranged in the form of equation~\eqref{rud7} with the matrix $\mathbb{M}$ having first column $\mathbb{M}_{j1}=\frac{1}{\sqrt5}$ for $j=1,\cdots, 5$, and the other columns 
\begin{align}
\mathbb{M}_{j2}{=}\!\begin{bmatrix}
 \frac{4 \mathcal{K}^3-2 \mathcal{K}^2-2 \mathcal{K}+1}{\sqrt{2} \left(\mathcal{K}^4-3 \mathcal{K}^2+1\right)} \\
 \frac{2 \mathcal{K}^4-5 \mathcal{K}^3+3 \mathcal{K}-1}{\sqrt{2} \left(\mathcal{K}^4-3 \mathcal{K}^2+1\right)} \\
 \frac{(1-2 \mathcal{K}) \mathcal{K}}{\sqrt{2} \left(\mathcal{K}^2-\mathcal{K}-1\right)} \\
 \frac{\mathcal{K}^2 \left(-2 \mathcal{K}^2+3 \mathcal{K}-1\right)}{\sqrt{2} \left(\mathcal{K}^4-3 \mathcal{K}^2+1\right)} \\
 \frac{\mathcal{K}^3 (2 \mathcal{K}-1)}{\sqrt{2} \left(\mathcal{K}^4-3 \mathcal{K}^2+1\right)} \\
\end{bmatrix}\!\!,\;\mathbb{M}_{ j3}{=}\!\begin{bmatrix}
 \frac{4 \mathcal{K}^4+2 \mathcal{K}^3-6 \mathcal{K}^2+1}{\sqrt{6} \left(\mathcal{K}^4-3 \mathcal{K}^2+1\right)} \\
 \frac{-2 \mathcal{K}^4+\mathcal{K}^3+2 \mathcal{K}^2-3 \mathcal{K}+1}{\sqrt{6} \left(\mathcal{K}^4-3 \mathcal{K}^2+1\right)} \\
 \frac{-2 \mathcal{K}^2-3 \mathcal{K}+2}{\sqrt{6} \left(\mathcal{K}^2-\mathcal{K}-1\right)} \\
 -\frac{\mathcal{K} \left(2 \mathcal{K}^3+\mathcal{K}^2-5 \mathcal{K}+2\right)}{\sqrt{6} \left(\mathcal{K}^4-3 \mathcal{K}^2+1\right)} \\
 \frac{\mathcal{K}^2 \left(2 \mathcal{K}^2+3 \mathcal{K}-2\right)}{\sqrt{6} \left(\mathcal{K}^4-3 \mathcal{K}^2+1\right)} 
\end{bmatrix}\!\!,\nonumber
\end{align}

\begin{align}
\mathbb{M}_{j4}{=}\!\begin{bmatrix}
 \frac{(1-2 \mathcal{K}) \mathcal{K}^3}{\sqrt{2} \left(\mathcal{K}^4-3 \mathcal{K}^2+1\right)} \\
 \frac{\mathcal{K}^2 \left(2 \mathcal{K}^2-3 \mathcal{K}+1\right)}{\sqrt{2} \left(\mathcal{K}^4-3 \mathcal{K}^2+1\right)} \\
 \frac{\mathcal{K} (2 \mathcal{K}-1)}{\sqrt{2} \left(\mathcal{K}^2-\mathcal{K}-1\right)} \\
 \frac{-2 \mathcal{K}^4+5 \mathcal{K}^3-3 \mathcal{K}+1}{\sqrt{2} \left(\mathcal{K}^4-3 \mathcal{K}^2+1\right)} \\
 \frac{-4 \mathcal{K}^3+2 \mathcal{K}^2+2 \mathcal{K}-1}{\sqrt{2} \left(\mathcal{K}^4-3 \mathcal{K}^2+1\right)} \\
\end{bmatrix}\!\!,\mathbb{M}_{j5}{=}\!\begin{bmatrix}
 \frac{-2 \mathcal{K}^4+5 \mathcal{K}^3+6 \mathcal{K}^2-2}{\sqrt{30} \left(\mathcal{K}^4-3 \mathcal{K}^2+1\right)} \\
 -\frac{2 \mathcal{K}^4+5 \mathcal{K}^3-11 \mathcal{K}^2+2}{\sqrt{30} \left(\mathcal{K}^4-3 \mathcal{K}^2+1\right)} \\
 \frac{-2 \mathcal{K}^2-3 \mathcal{K}+2}{\sqrt{30} \left(\mathcal{K}^2-\mathcal{K}-1\right)} \\
 \frac{-2 \mathcal{K}^4+5 \mathcal{K}^3-4 \mathcal{K}^2-5 \mathcal{K}+3}{\sqrt{30} \left(\mathcal{K}^4-3 \mathcal{K}^2+1\right)} \\
 \frac{8 \mathcal{K}^4-14 \mathcal{K}^2+3}{\sqrt{30} \left(\mathcal{K}^4-3 \mathcal{K}^2+1\right)} \\
\end{bmatrix}\!\!,\nonumber
\end{align}
where the frequency dependent factor $\mathcal{K}$ is defined as $1/\mathcal{K}=2-i\omega\tau$.

\section{Charge averaging}\label{ca}
To arrive from equation~\eqref{rud3} to the backscattering Hamiltonian~\eqref{rud9}, we first define the coefficients appearing in equation~\eqref{rud3} by
\begin{align}
U_\alpha=\lambda_1 D\mathsf{U}_\alpha,\;\;\;V_\alpha=\lambda_2 D^4\mathsf{V}_\alpha,
\end{align}
where $D$ is the usual band cutoff. The cosine terms are then averaged by integrating out the high energy (${\rm he}$) part, gapped modes, of the equation~\eqref{rud7} using the scattering matrix $\mathbb{M}$ given in appendix~\ref{a1}. Such a procedure gives $e^{ {-\frac{1}{2}}\left<\Phi_{1,5}^2\right>_{\text{he}}}=\mathcal{P}_1$, $e^{ {-\frac{1}{2}}\left<\Phi_{2, 4}^2\right>_{\text{he}}}=\mathcal{P}_2$ and $e^{ {-\frac{1}{2}}\left<\Phi_{3}^2\right>_{\text{he}}}=\mathcal{P}_3$ for the functions $\mathcal{P}_\alpha$ defined by
\begin{align}
\mathcal{P}_\alpha&=\mathcal{Y}_\alpha\!\left(\frac{5^{3/16} e^{\gamma }}{ D \tau }\right)^{\frac{4}{5}}\!\!\!\!,\;
\mathcal{Y}_1{ =}\left(72 \sqrt{5}{+}161\right)^{{-}\frac{1}{8 \sqrt{5}}}\simeq 0.724,\nonumber
\end{align}
\begin{align}
\mathcal{Y}_2 &=\frac{3 \coth ^{-1}\left(\frac{3}{\sqrt{5}}\right)}{4 \sqrt{5}}\simeq 1.381,\;\mathcal{Y}_3 =\left(\frac{5}{16}\right)^{4/5}\simeq 0.394.\nonumber
\end{align}
The parameters $\mathcal{C}_\alpha$ and $\mathcal{D}_\alpha$ in equation~\eqref{rud10} are connected such that $\mathcal{C}_\alpha=\lambda_1 \mathcal{Y}_\alpha$ and $\mathcal{D}_\alpha=\lambda_2\mathcal{Y}^4_\alpha$.

\end{document}